\documentclass[letter,
               keeplastbox,   
               ]{jacow}
\usepackage{pdfpages,multirow,ragged2e} %
\makeatletter%
	\ifboolexpr{bool{xetex}}
	 {\renewcommand{\Gin@extensions}{.pdf,%
	                    .png,.jpg,.bmp,.pict,.tif,.psd,.mac,.sga,.tga,.gif,%
	                    .eps,.ps,%
	                    }}{}
\makeatother

\ifboolexpr{bool{xetex} or bool{luatex}} 
 {}                                      
 {\usepackage[utf8]{inputenc}}           

\usepackage[USenglish]{babel}

\ifboolexpr{bool{jacowbiblatex}}%
 {%
  \addbibresource{jacow-test.bib}
  \addbibresource{biblatex-examples.bib}
 }{}
\begin{document}
\title{Low phase noise master oscillator generation and distribution for \NoCaseChange{ALS} and \NoCaseChange{ALS-U}}


\author{M. Betz\thanks{mbetz@lbl.gov}, Q. Du, B. Flugstad, K. Baptiste, M. Vinco\\
Lawrence Berkeley National Laboratory, Berkeley, USA}

\maketitle

\begin{abstract}
The coax based MO distribution system in the ALS is going to be replaced by a
modernized, lower phase noise and more interference tolerant version,
ready to support ALS-U operation. System aspects are shown and several
commercial analog and digital optical transceiver modules are compared for
their suitability in this application. Furthermore, recent phase noise
optimizing efforts in the ALS RF system are discussed and several prototypes
for a custom built, low phase noise, frequency adjustable master oscillator
around 500 MHz are shown.
\end{abstract}

%

\section{MO distribution}
The Advanced Light Source (ALS) is clocked by a single Master Oscillator (MO),
running at $f_1 \approx 499.6$~MHz and distributed to several clients
across the accelerator complex through phase stable coax
(Andrew LDF4-50A Heliax).
A 12 x distribution chassis, based on 10~W RF power amplifier
modules (MHW709-3) and a custom level control loop,
provide enough power to drive the long coax cables \cite{ref:als_mo}.

\begin{figure}[!htb]
   \centering
   \includegraphics*[width=.7\columnwidth]{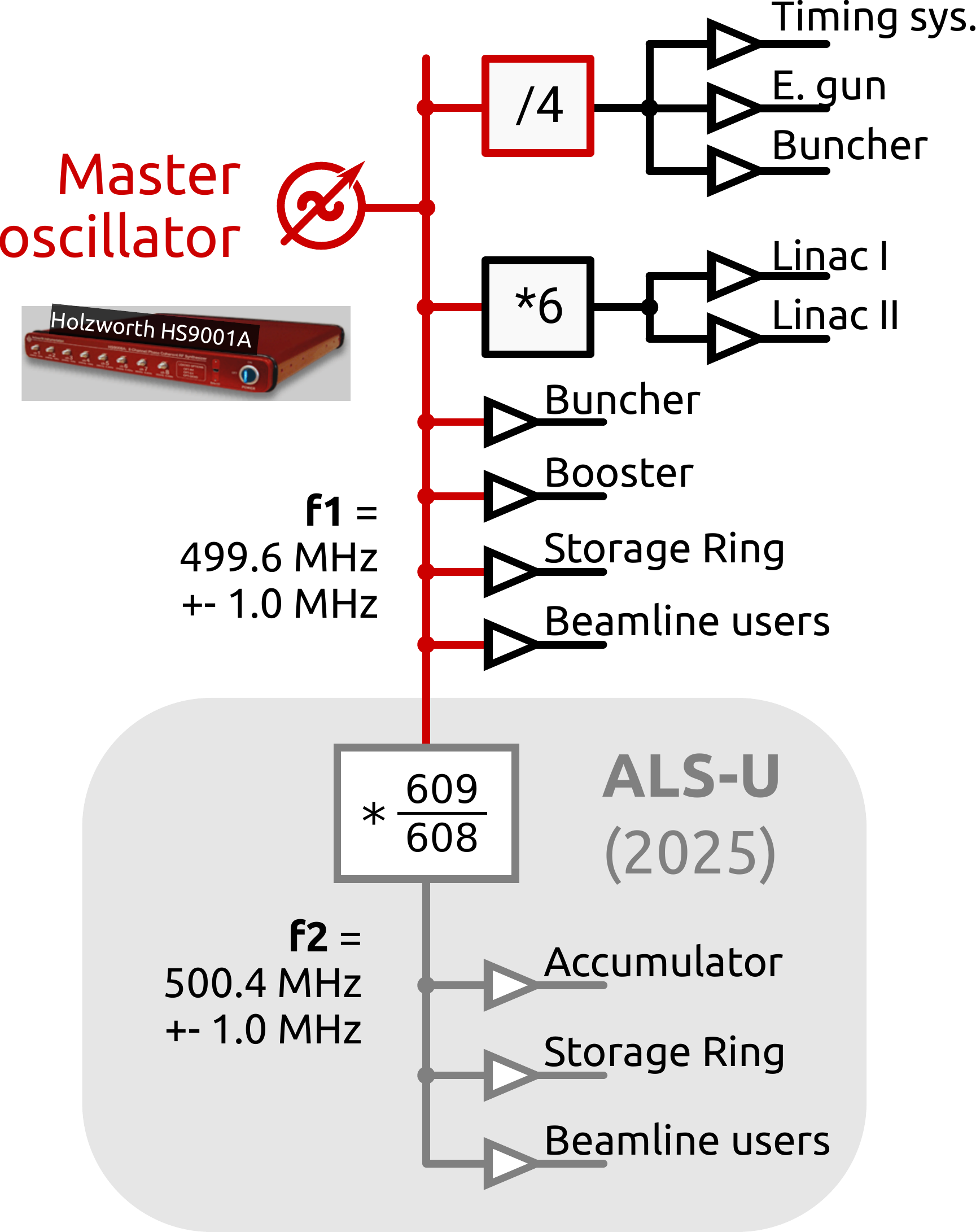}
   \caption{Layout of the ALS MO distribution system.
   The additional $f_2$ endpoints for the future ALS-U ring
   are shown in the grayed out box at the bottom.}
   \label{fig:tree}
\end{figure}

Additionally, a divide by 4 clock is generated by an AD9513 evaluation board
in the same rack and distributed, without additional amplifiers, to
the nearby timing, gun and buncher LLRF systems.
The S-band linac is clocked by the $f_1$ distribution and derives its
$f_1 \cdot 6 \approx 3$~GHz RF clock through a local multiplier. The clock
tree has been illustrated in Fig.~\ref{fig:tree}.

The $f_1$ distribution system has been operational with high reliability since its
installation in 1989, when the ALS was built.
Nonetheless, many of its parts have become obsolete.
Additionally, more beamline users are sensitive to timing and require
a low jitter frequency reference. The 12 channel limit has been exhausted.
Workarounds with RF power splitters have caused phase shift problems in the past,
when a port was not properly terminated.

\begin{figure}[!htb]
   \centering
   \includegraphics*[width=.7\columnwidth]{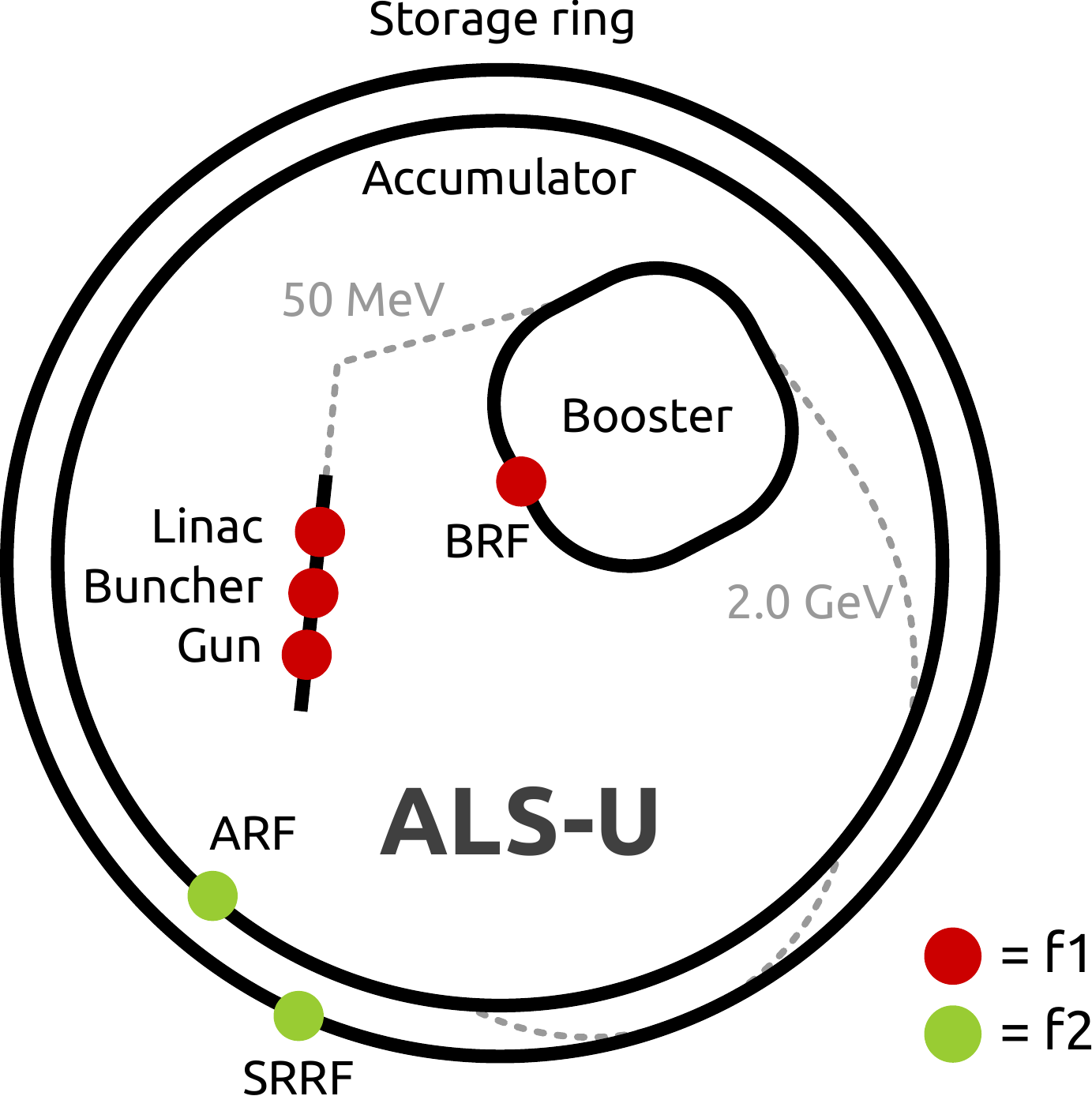}
   \caption{
   RF systems of the ALS-U. Color indicates reference frequency.
   In comparison, the ALS does not have an accumulator ring and SRRF
   is clocked by $f_1$.
   }
   \label{fig:ALS-U}
\end{figure}

The current plans for the ALS-upgrade project (ALS-U) foresee a new
storage ring based on a multi-bend achromat lattice.
Its circumference is shorter by $\frac{1}{2}$ RF wavelength ($\approx 30$~cm),
increasing its RF frequency to $f_2 = f_1 \frac{609}{608}$.
The choice was made to operate the accumulator and storage ring at $f_2$ while
keeping the injector and booster ring running at $f_1$ \cite{ref:alsu_f2}.
This avoids the need of
re-aligning or, in case of the linac, re-building these accelerators.
The ALS-U layout, as currently planned, is shown in Fig.~\ref{fig:ALS-U}.
The current MO distribution system is not flexible enough to support
2 RF frequencies.

\begin{figure}[!htb]
   \centering
   \includegraphics*[width=\columnwidth]{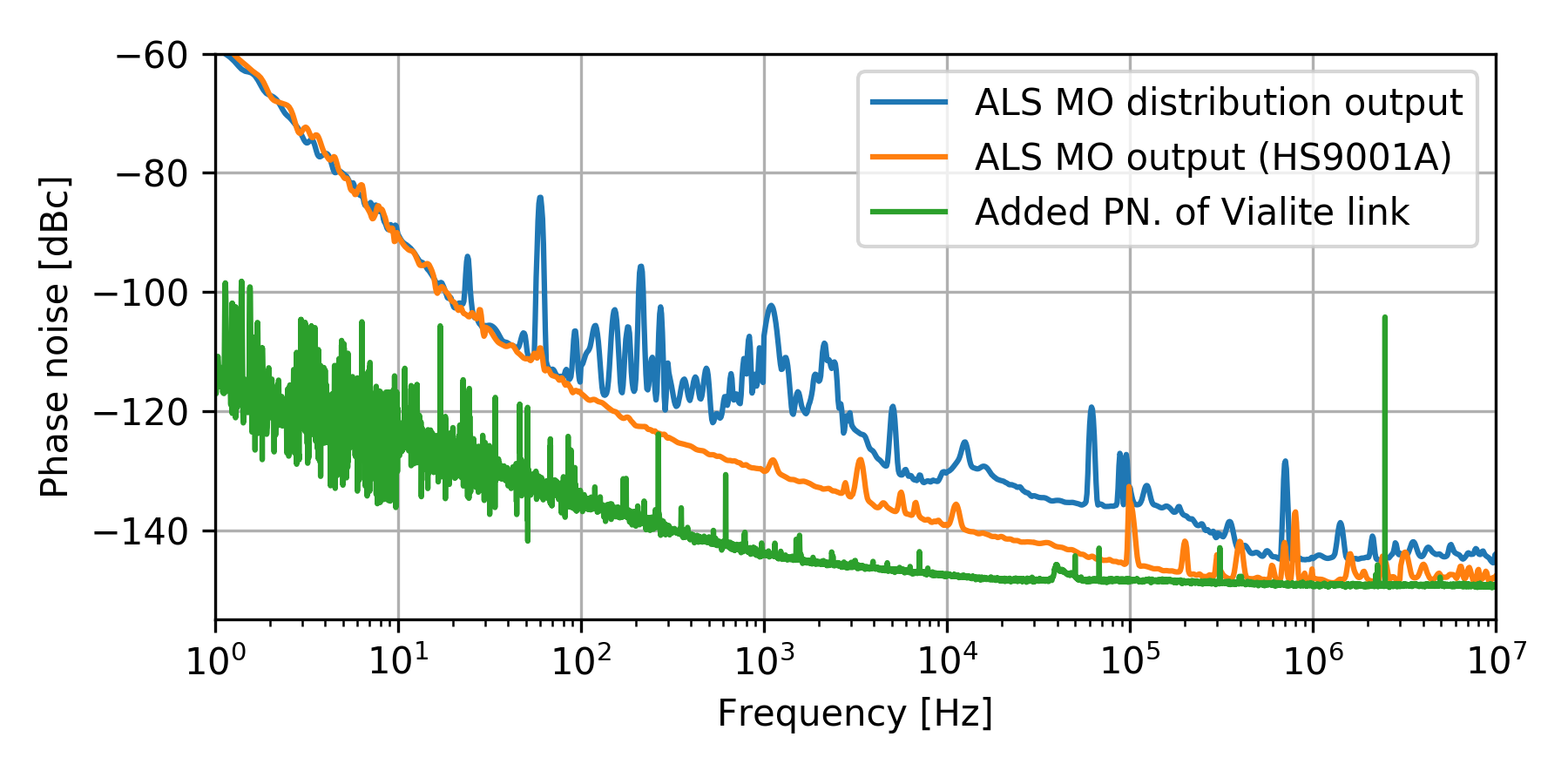}
   \caption{
    Orange and blue trace: Measurement of Phase noise at the in- and output of the
    ALS MO distribution chassis. Green: Measured added phase noise of the
    proposed Vialite RF over fiber link with optical splitter.
   }
   \label{fig:pn_dist}
\end{figure}

Last but not least, the distribution chassis has been found to add a
significant amount of phase noise to the signal as shown in Fig.~\ref{fig:pn_dist}.

For these reasons the ALS MO distribution system will be upgraded in the near future.
An optical fiber based solution has been chosen due to the many advantages over coax based systems. This includes excellent EMI and crosstalk rejection, inherent galvanic isolation, significantly lower transmission loss, higher
fan-out capability and the absence of powerful microwave distribution
amplifiers with their potential power dissipation and signal quality issues.

The commercial RF over Fiber system from Vialite has been chosen due to its
very good phase noise performance. The added phase noise of an optical
transmitter, a 16 x active optical splitter and an optical receiver,
shown in Fig.~\ref{fig:pn_dist} (green trace),
is negligible compared to the MO phase noise (orange trace).

Furthermore, the Vialite system provides useful reliability features,
like redundant power supplies, a chassis with hot-swappable
transceiver modules, blind mate connectors and the ability to monitor the
system over ethernet.

The dual frequency scheme of ALS-U is supported and the system
offers ultimate flexibility as each transmitter / receiver
pair can operate on a different frequency.

RF signals within a usable bandwidth of 1~GHz are amplitude modulated on an optical carrier.
Amplification and fan-out happens in the optical domain by a commercial `lossless splitter`,
which provides extremely high isolation of the output ports from each other.
Hence a bad RF termination cannot lead to phase errors on other outputs.

\begin{figure}[!htb]
   \centering
   \includegraphics*[width=\columnwidth]{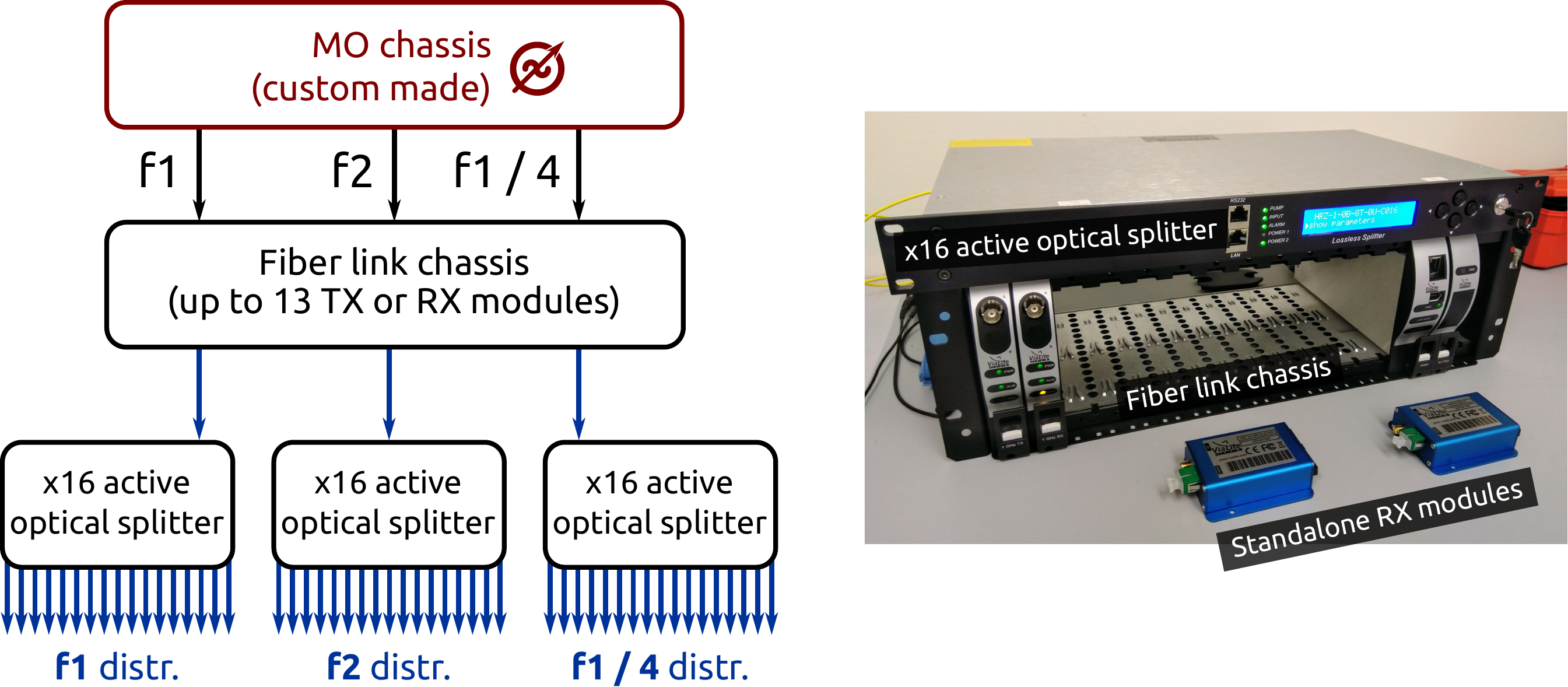}
   \caption{
    Left: Possible layout of the new fiber based MO distribution system for the ALS.
    Right: Commercial components which will be used.
   }
   \label{fig:dist_sketch}
\end{figure}

Each distribution end-point will be equipped with a standalone optical receiver module
and, if necessary, with an additional microwave power amplifier and
bandpass filter to reduce out of band spurious signals.
The components which will be used and a possible distribution scenario
are shown in Fig.~\ref{fig:dist_sketch}.

\section{New MO for ALS and ALS-U}
The requirements for a new ALS MO include lowest possible phase noise and low spurious coherent signals (spurs) in the 10~Hz to 10~kHz offset range. For slow orbit feedback, frequency set-point needs to be
adjustable in 1~mHz increments around $f_1 \pm 100$~kHz  with $\approx 1$~Hz update rate.
Output phase and amplitude needs to be continuous during these adjustments.

The ALS operational MO was switched from HP8644b
to Holzworth HS9001A in January 2019. This reduced the integrated phase noise
within 1~Hz to 1~MHz from 11.5~ps to 0.5~ps.
For frequency tuning, the old MO relied on a workaround of using an external
DC control voltage on its FM modulation input -- which has caused scaling and
out-of-range problems in the past.
The new MO can update its digital frequency set-point without phase-glitch
and does not require this workaround, simplifying ALS operation.

\begin{figure}[!htb]
   \centering
   \includegraphics*[width=\columnwidth]{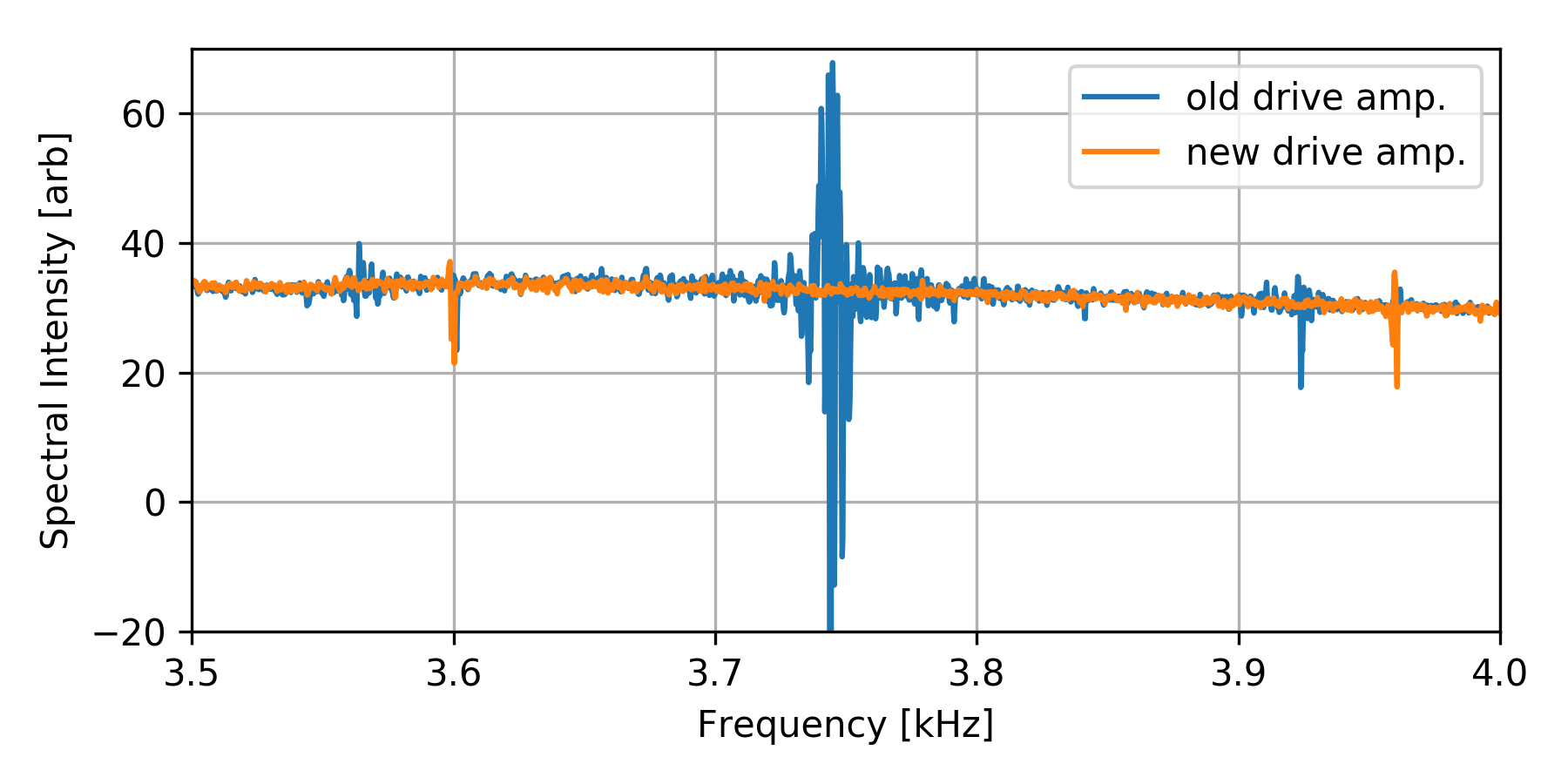}
   \caption{
    The spectral infrared beamline measurement shows noisy data before (blue trace) and
    clean data after (orange trace) a new klystron drive amplifier was installed in the
    storage ring RF system. The old drive amp caused a spur at 3.75 kHz offset of
    $\approx -100$~dBc magnitude.
   }
   \label{fig:infrared}
\end{figure}

Previous experience has shown, the ALS infrared beamline is sensitive to spurs in the MO phase noise spectrum \cite{ref:bechtel}, particularly in the 10~Hz to 10~kHz carrier offset range. These spurs transfer trough the LLRF system onto the beam, modulating the bunch arrival time.
A good example is a spur around 3.75~kHz of around -100~dBc magnitude,
which was tracked down to a switch mode power supply within
the SRRF klystron drive amplifiers.
The spur was clearly visible in the beam spectrum of the longitudinal feedback
system and in the spectral measurement results of the ALS infrared beamline,
as shown in Fig.~\ref{fig:infrared}.
The spur magnitude was reduced by > 20~dB after replacing the drive amplifiers in May 2018, improving the signal to noise ratio during infrared beamline measurements.

\begin{figure*}[!htb]
   \centering
   \includegraphics*[width=\textwidth]{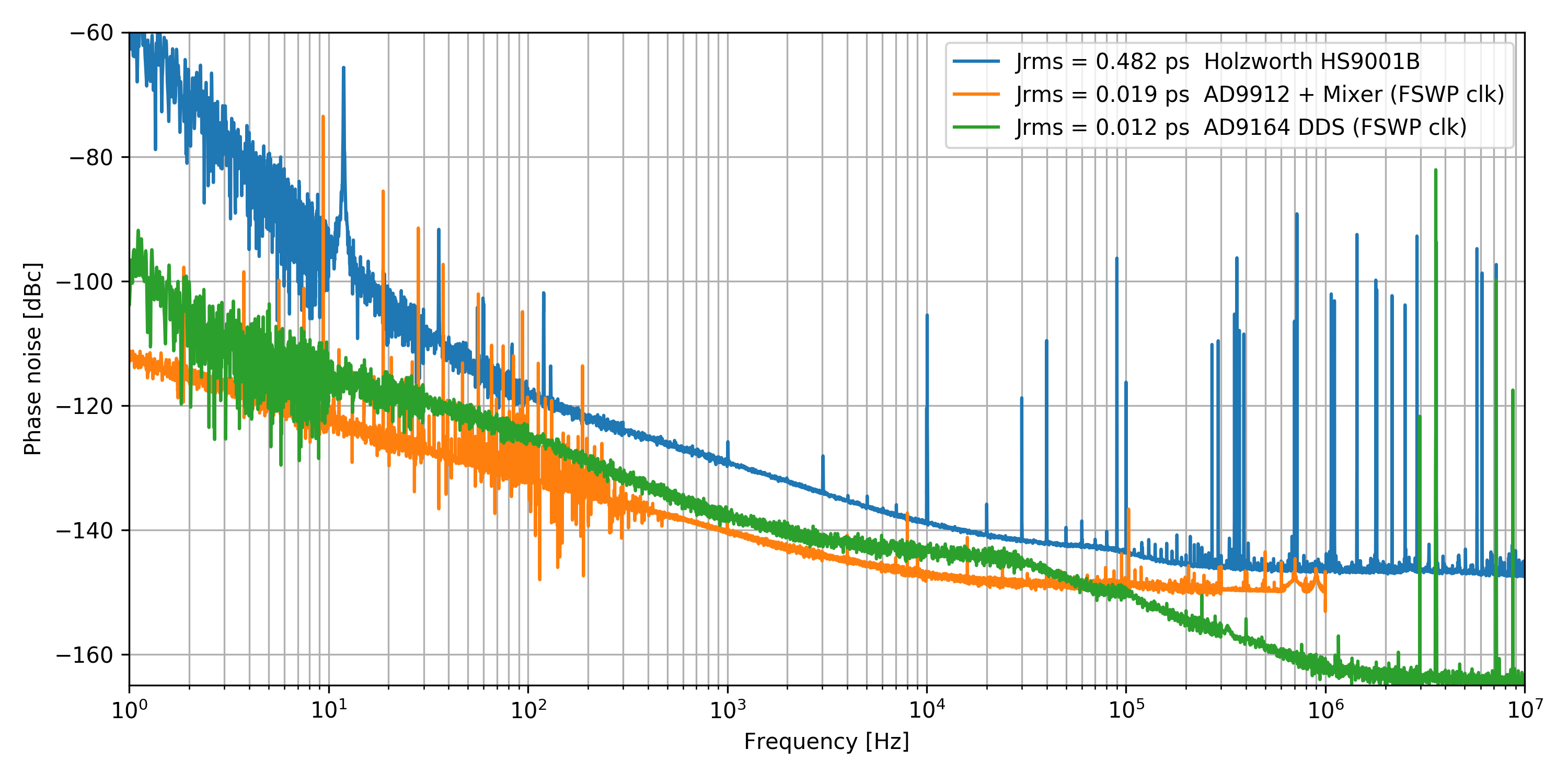}
   \caption{
    Measured phase noise and spurs of a Holzworth HS9001B (blue), which is the
    current MO in the ALS, and the two custom made solutions described in
    the text (orange and green). Carrier at 499.65~MHz for all 3 traces.
   }
   \label{fig:pn_homebrew}
\end{figure*}

Keeping this sensitivity to spurs in mind, the blue trace in
Fig.~\ref{fig:pn_homebrew} indicates some room for improvement for the HS9001.
As no other commercial instrument with better broadband phase noise, better spurious performance and phase continuous frequency adjustment capability could be found, two ideas for a custom built MO were further investigated.

\subsection{AD9912 DDS + Mixer}
\label{sec:hb_1}
\begin{figure}[!htb]
   \centering
   \includegraphics*[width=0.9\columnwidth]{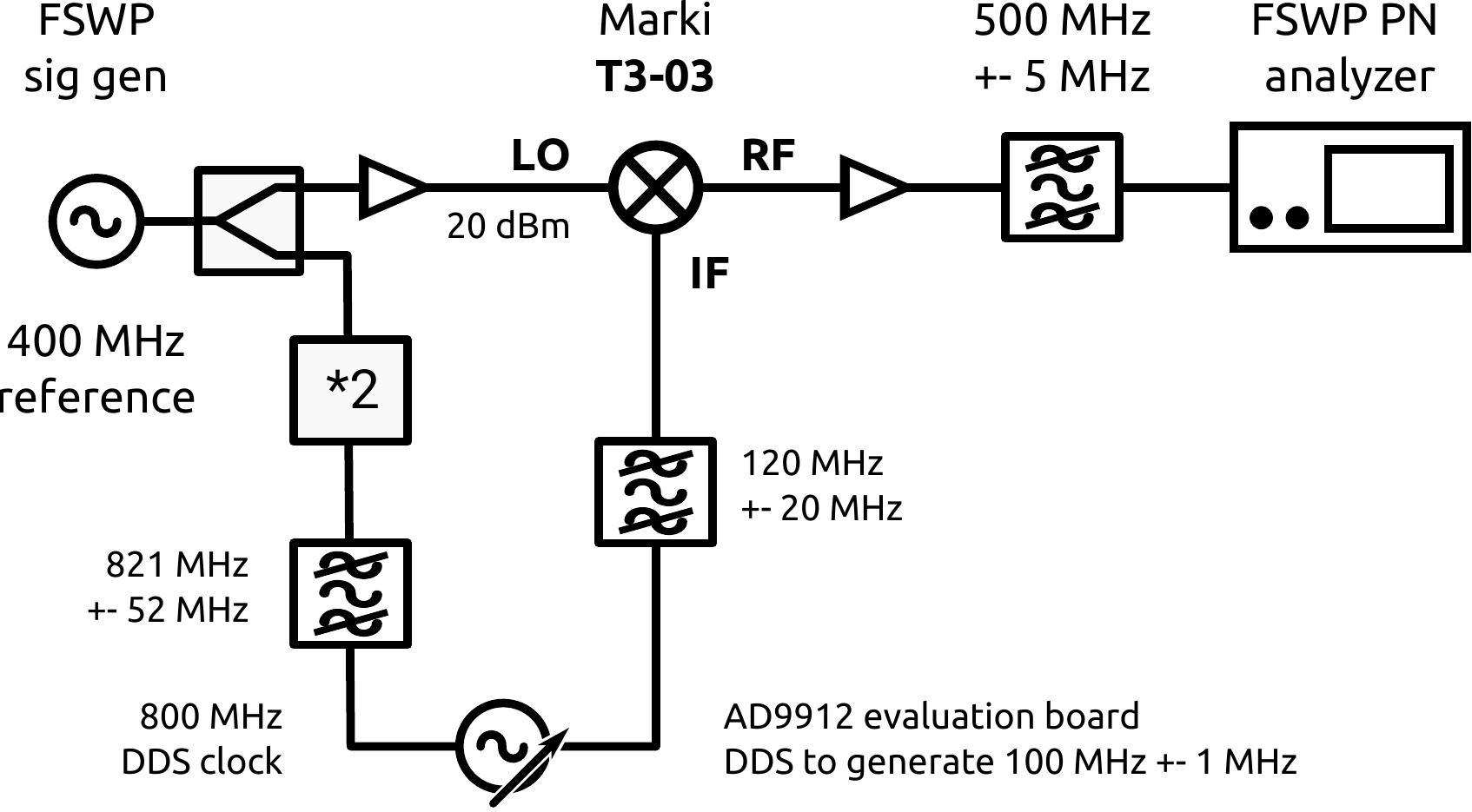}
   \caption{
    Setup for synthesizing $f_1$ by up-converting the variable frequency output of a AD9912 DDS.
   }
   \label{fig:hb_1}
\end{figure}

A clean fixed 400 MHz frequency reference is split in two channels.
One is doubled and used as clock for a AD9912 Direct Digital Synthesis (DDS) chip,
generating an adjustable frequency of $\approx$~100~MHz. The other one is used as Local Oscillator for a mixer, to up-convert the DDS output to an adjustable ~500 MHz MO output. The setup for phase noise and spur measurements is shown in Fig.~\ref{fig:hb_1}. The internal signal generator
of the FSWP signal source analyzer was used as 400 MHz frequency reference. The FSWP rejects the noise of its internal source, hence this is an additive phase noise measurement. For the operational system, a high quality fixed frequency OCXO needs to be used, which will add its own noise to the budget.

The measured phase noise is shown in Fig.~\ref{fig:pn_homebrew} (orange trace).
While phase noise performance is on average the best of the MO sources which have
been considered so far, the setup suffers from significant spurs in the sensitive frequency range. These spurs move with frequency set-point and are hence hard to control. They originate from the AD9912 output and are inherent to its limited 14~bit DAC resolution \cite{ref:dds_spurs,ref:dds_spurs_rubi}.

\subsection{Direct synthesis with AD9164 DAC}
\label{sec:hb_2}
\begin{figure}[!htb]
   \centering
   \includegraphics*[width=0.8\columnwidth]{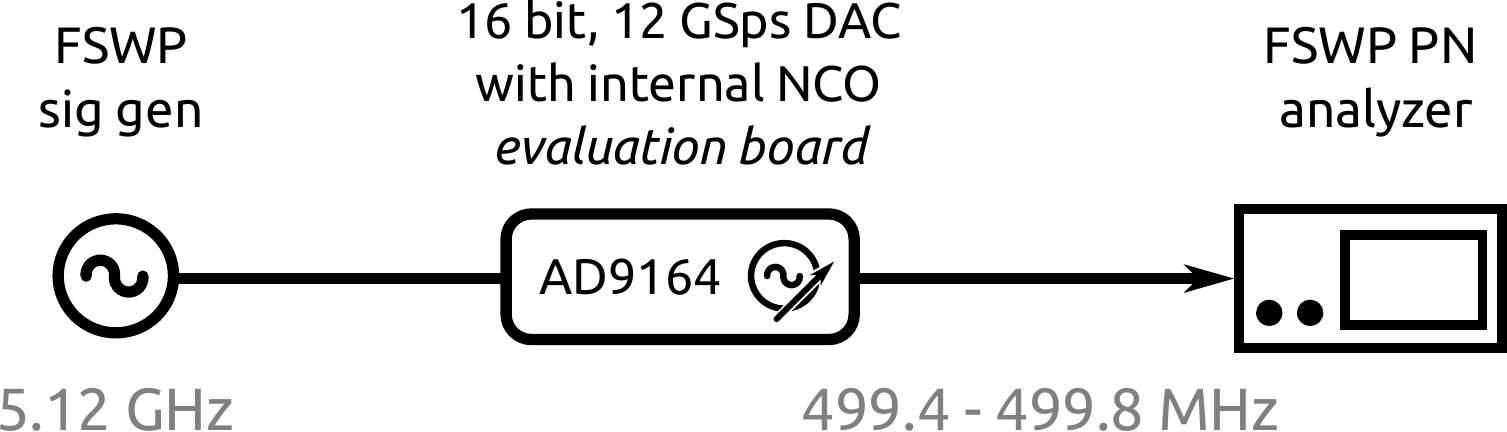}
   \caption{
    Setup for evaluating phase noise and spurs of the AD9164 DAC, which can potentially be used to directly synthesize $f_1$.
   }
   \label{fig:hb_2}
\end{figure}

\begin{figure*}[!htb]
   \centering
   \includegraphics*[width=\textwidth]{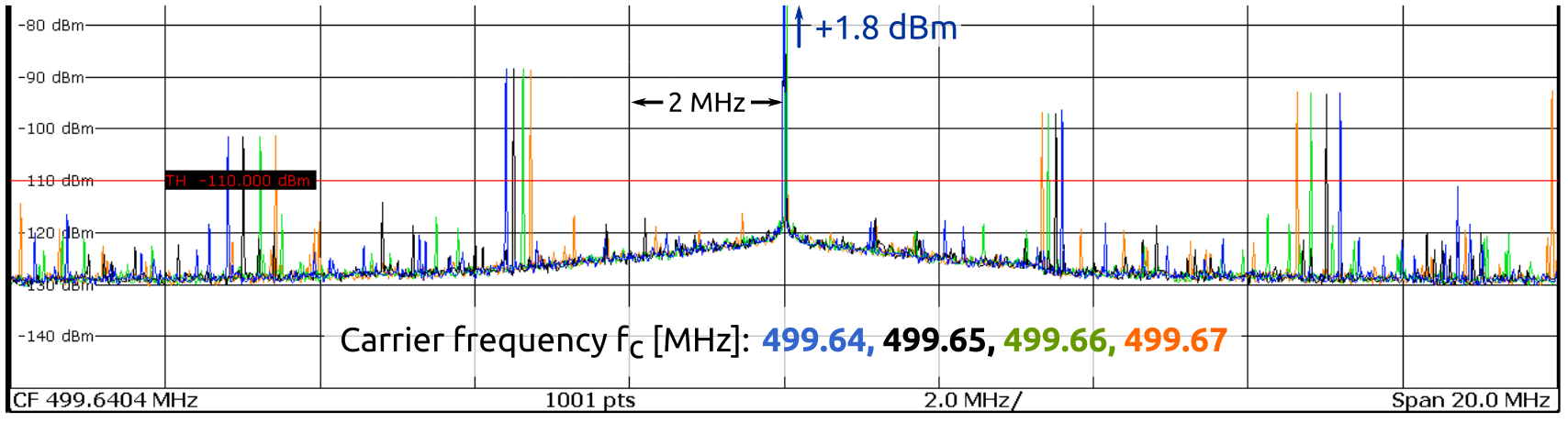}
   \caption{
    Spectrum analyzer measurement of AD9164 SFDR.
    The DAC was clocked at $f_\mathrm{DAC} = 5.12$~GHz and
    carrier frequency was kept within 499.4~MHz~$< f_C <$~499.8~MHz. No
    spurs larger than -117 dBc within a $\pm$2~MHz window around the carrier were observed.
    This window was chosen to allow mitigation of the remaining spurs with a narrow bandpass filter at the DAC output.
   }
   \label{fig:ad9164_spurs}
\end{figure*}
To improve upon spurious performance, a DDS with higher resolution than the
AD9912 is needed. The AD9164, a modern DAC chip,
achieves 16~bit resolution at up to 12~GSps and is
hence capable of directly synthesizing $f_1$ without the need for up-conversion.
It contains internal DDS functionality which was used to generate a test-tone
at 499.65~MHz to evaluate its phase noise performance. The sampling clock $f_\mathrm{DAC} = 5.12$~GHz was chosen for good spur performance at ALS carrier frequencies and
derived from the FSWP internal signal generator as shown in Fig.~\ref{fig:hb_2}.

The green trace in Fig.~\ref{fig:pn_homebrew} shows the AD9164 added phase noise.
While the average noise floor is slightly higher than for the AD9912 setup, there are
no significant spurious signals within $\pm2$~MHz of the carrier.
This was further verified with multiple spectrum analyzer measurements at different carrier frequency ($f_C$) set-points as shown in Fig.~\ref{fig:ad9164_spurs},
confirming a Spurious Free Dynamic Range (SFDR) of > 117~dB.
Both measurements show some larger spurs in the -90~dBc range at $\approx 3$~MHz offset.
These are far enough from the carrier to use a narrow bandpass filter at the AD9164 output for mitigation.

A dual output version (AD9174) of this DAC is available, which can generate the additional $f_2$ for ALS-U operation.
When testing the internal DDS functionality of the AD9174, it was found that updating the Frequency Tuning Word
causes the output phase to jump to random values. Hence the DDS logic needs to be implemented in an external FPGA.
This adds complexity but also allows for ultimate flexibility and makes it possible to implement advanced features like DDS
spur suppression through dithering or destructive interference.
The Xilinx VC707 evaluation board was chosen for this purpose,
it can drive up to 2 x AD9174-FMC-EBZ DAC boards and hence can provide up to
4 independent RF output channels.

Synthesizing the 2 rationally related frequencies $f_1$ and $f_2$ can be achieved with 2 independent phase accumulators, as long as the Frequency Tuning Word is updated synchronously within the same DSP clock cycle, to keep the 2 phases in sync. To achieve precise frequency ratios other than $N / 2^M$, a modulo logic will be required, which also needs to be updated synchronously.
Further work is necessary to synchronize the 2 phase accumulators at regular intervals, which would be a robust way of avoiding the accumulation of phase errors between the 2 outputs.

\section{CONCLUSION}
A new MO distribution system based on commercial RF over fiber technology has been proposed and evaluated for usage in the ALS. Its active optical splitters provide significant advantages compared to the traditional approach of amplifying and splitting the RF signal in the electrical domain.

Two custom built candidates for a cleaner ALS MO have been investigated, both having the potential to improve over the current MO (HS9001A) performance in terms of phase noise and spurs.

The approach based on the fast DAC (AD9164) is preferred due to its minimal additional hardware requirements. Carrying out the frequency synthesis in a FPGA provides ultimate flexibility. Even though the average phase noise noise-floor is slightly higher compared to the AD9912 based setup, its spur performance is much improved. Carefully choosing the DAC sampling rate, an essentially spur-free window has been found $\pm2$~MHz around the carrier, when operating at typical frequencies for the ALS.

More work is needed designing and implementing the frequency synthesis logic and the JESD204B interface on the FPGA.

\section{ACKNOWLEDGMENTS}
The authors are grateful for practical hints and assistance from L. Doolittle, G. Huang, S. F. Paiagua, C. S. Pareja and V. Vytla.
This work was supported by the Director, Office of Science, Office of Basic Energy Sciences, of the U.S. Department of Energy under Contract No. DE-AC02-05CH11231.

%
%
\ifboolexpr{bool{jacowbiblatex}}%
	{\printbibliography}%
	{%

	
} 

%
%


\end{document}